\documentclass[twocolumn,pra,aps,amsmath,amssymb,nofootinbib]{revtex4-2}
\usepackage{graphics}
\usepackage{graphicx}
\usepackage{dcolumn}
\usepackage{bm}
\usepackage{mathrsfs}
\usepackage{pstricks}
\usepackage{color}
\usepackage{soul}
\usepackage{float}
\usepackage{slashed}
\usepackage{amsmath}
\usepackage{epsfig}
\usepackage{amsfonts}
\usepackage{amssymb}

\def\beq{\begin{equation}}
\def\eeq{\end{equation}}
\def\bea{\begin{eqnarray}}
\def\eea{\end{eqnarray}}

\begin{document}

\title{Time-dependent quantum harmonic oscillator: a continuous route from adiabatic to sudden changes}
\author{D. Mart\'inez-Tibaduiza}
\email{Correspondence to: danielmartinezt@gmail.com}
\affiliation{Instituto de F\'isica, Universidade Federal Fluminense, \\ 
 Avenida Litor\^{a}nea, 24210-346 Niteroi, RJ, Brazil}
\author{L.~Pires}
\affiliation{Institut de Science et d'Ing\'enierie Supramol\'eculaires, CNRS, Universit\'e de Strasbourg, \\ 
UMR 7006, F-67000 Strasbourg, France}
\author{C. Farina}
\affiliation{Instituto de F\'isica, Universidade Federal do Rio de Janeiro, \\ 
21941-972 Rio de Janeiro, RJ, Brazil}
%

\begin{abstract}

In this work, we provide an answer to the question: how sudden or adiabatic is a 
change in the frequency of a quantum harmonic oscillator (HO)?
To do this, we investigate the behavior of a HO, initially in its fundamental state, 
by making a frequency transition that we can control how fast it occurs. 
The resulting state of the system is shown to be a vacuum squeezed state in two bases 
related by Bogoliubov transformations. We characterize the time evolution of the squeezing parameter in both bases 
and discuss its relation with adiabaticity by changing the transition rate
from sudden to adiabatic. 
Finally, we obtain an analytical approximate expression that relates squeezing to the 
transition rate as well as the initial and final frequencies. 
Our results shed some light on subtleties and common inaccuracies 
in the literature related to the interpretation of the adiabatic theorem for this system.

\end{abstract}



\maketitle

\section{Introduction}\label{intro}

The harmonic oscillator (HO) is undoubtedly one of the most important systems in physics since 
it can be used to model a great variety of physical situations both in  classical and  quantum contexts. 
In the former case, almost all movements of physical systems with small amplitude around a stable equilibrium configuration 
are described by harmonic motions which explain, among many other things, 
why the classical theory of dispersion in dielectrics works so well \cite{Zangwill-2013,Jackson-2007}. 
In the quantum context, we can mention many examples, from quantum optics, where for many purposes the 
quantized modes of the electromagnetic field behave like quantum HOs \cite{Loudon-2000}, 
to quantum chemistry, where the main characteristics of the London-van der Waals forces 
can be understood by considering two atoms as fluctuating dipoles modeled by 
oscillating charges interacting through their instantaneous dipole fields \cite{Milonni-2013}.

A more general system may be a time-dependent harmonic oscillator  (TDHO), \textit{i.e.}, 
a HO whose parameters: mass, frequency or both, are time-dependent. 
This system has been very well studied since the Husimi's first solution in 1953 \cite{Husimi-1953}, 
and it is a 
natural scenario to study 
the important squeezed states, appearing in a variety of branches of physics such as quantum optics 
\cite{WALLS-1983, LOUDON-1987, WU-1987, TEICH-1989, GLAUBER-1991, PERINA-1991, DODONOV-2002, DODONOV-2003, SCHNABEL-2017}, 
gravitational interferometry 
\cite{GRISHCHUK-1983, GEA-BANACLOCHE-1987, HARMS-2003, GROTE-2013, DEMKOWICZ-DOBRZANSKI-2013, DWYER-2013, AASI-2013, ABBOTT-2016}, 
cosmology \cite{GRISHCHUK-1990, GRISHCHUK-1993, ALBRECHT-1994, HU-1994, EINHORN-2003,KIEFER-2007}, 
metrology \cite{VAHLBRUCH-2007, GIOVANNETTI-2011}, telecommunications \cite{SLAVIK-2010},  
analogue models to the dynamical Casimir effect 
\cite{DODONOV-2005,JOHANSSON-2009,JOHANSSON-2010,DODONOV-2010,WILSON-2011,FUJII-2011,LAHTEENMAKI-2013,FELICETTI-2014} 
and spin states \cite{Masahito-1993,Vuletic-2010,Hosten-2016}, relevant in optical clocks \cite{Chou-2010}. 
The main property of these states is to reduce the value of one of the quadrature variances 
(the variance of the orthogonal quadrature is increased accordingly) 
in relation to coherent states, 
which already saturate the Heisenberg relation \cite{BO-STURE-1985, WODKIEWICZ-1985, GAZEAU-2009}. 
This property is known as squeezing.   
Another context where the study of the TDHO is relevant is shortcuts to adiabaticity. There, 
it is used in the determination of optimal non-adiabatic protocols connecting two equilibrium states 
with different frequencies, in such a way that the populations of the final and initial states are identical,   
a result naturally obtained through a perfectly adiabatic transformation 
\cite{Chen-2010, schaff2010fast, Del-Campo-2011, guery-2019}.
It is worth emphasizing that the study of the TDHO is also relevant in the context of quantum thermodynamics, 
where it can be considered as a working medium in a thermodynamical cycle \cite{Torrontegui_2013, Kosloff_2017}, 
with the frequency changes being the quantum equivalent to volumetric expansions and contractions of the medium. 
In the above topics, adiabaticity is a concept of fundamental importance.

Adiabaticity in the TDHO, and its relation with squeezing, has been extensively studied since its first solution. 
For instance, it is well-known that any non-adiabatic change in the parameters of a HO, initially in a coherent state, 
generates squeezing \cite{Husimi-1953, Graham-1987}. 
As an important case, a sudden change (or, simply a jump) in the frequency of a HO, 
has exact solution and it has been very well studied and characterized 
\cite{JANSZKY-1986, Rhodes-1989, C.F.LO-1990, CFLO-1991, Kumar-1991, JANSZKY-1992, JANSZKY-TE-1994, JANSZKY-1994, MOYA-2003, DMTibaBJP-2020}. 
On the other hand, adiabatic changes do not produce squeezing and
the so-called quantum adiabatic invariants of the system can be identified 
and used to solve the dynamics of the system 
\cite{Husimi-1953, LEWIS-1969, Popov-1969, Malkin-1970, Man'Ko-1970, DODONOV-2005, Boldt_2013}. 
Nevertheless, up to the authors' knowledge, in the literature there is not a full characterization of squeezing 
for a continuous interpolation between the sudden and adiabatic limits, 
enabling one to answer how sudden or adiabatic is a given change in the parameters of a HO. 


In this paper, we study numerically the time evolution of a HO 
whose time-dependent frequency can be adjusted in such a way that we can pass 
from an abrupt change to an adiabatic one continuously. 
Our starting point is the solution for a TDHO developed in Ref. \cite{DMTibaTDHO-2020}, 
valid for an arbitrary frequency function and any initial state. 
This solution is extremely convenient to do numerical calculations, 
and given in the initial basis, 
which diagonalizes the hamiltonian at $t=0$. 
We extend this solution to the instantaneous basis description, 
which diagonalizes the hamiltonian at any time, 
considering the initial state as the fundamental one.  
The aforementioned bases are related by Bogoliubov transformations, 
and we do explicit calculations for the change of basis 
using BCH-like relations established in Ref. \cite{DMTibaBCH-2020}. 
An important advantage of our results, is that 
the solution in the instantaneous basis remains 
suitable for direct numerical implementation. 
The frequency transition is modeled with an emblematic family of monotonic functions that 
asymptotically approach the initial and final values of the frequency 
(see Eq. (\ref{eq:frequency1}) and panel (a) of Fig. \ref{fig:FreqVsTime}),  
containing a rate control parameter. 
Using this model, we initially discuss the relation between squeezing and adiabaticity, 
clarifying some subtleties and common inaccuracies present in the literature 
related to the interpretation of the adiabatic theorem. 
Then, we use numerical analysis to 
fit a simple but very accurate analytic expression 
for the final squeezing in the HO 
as a function of the parameter that characterizes how fast the frequency changes 
(once the initial and final frequencies are given). 
Based on this expression we deduce 
the adiabatic limit 
and generate contour plots, 
providing a quick selection of the parameters to obtain a desired amount of squeezing. 
The analysis of these results reveals an outstanding difference in the squeezing production 
between enhancing or diminishing the frequency of a HO.

This paper is organized as follows. In Section \ref{IB} we present the mathematical survey for the time evolution 
of the system. In Section \ref{SvsA} we perform numerical calculations to analyze squeezing and adiabaticity in a frequency transition. 
In Section \ref{CINT} we present the continuous interpolation between sudden and adiabatic transitions. 
Section \ref{concl} is left for the conclusions and final remarks.

\section{Time evolution}\label{IB}

Consider a one-dimensional HO of unit mass with arbitrary time-dependent frequency, $\omega(t)$. Its hamiltonian is given by 
\begin{equation}
\hat{H}(t)=\frac{1}{2}\hat{p}^{2}+\frac{1}{2}\omega^{2}(t)\hat{q}^2\, ,
\label{eq:HOVF}
\end{equation}
 where $\hat{p}$ and $\hat{q}$ are the usual momentum and position operators in the Schr\"odinger picture.
 Let us call $\hat{H}(t)$ the \textit{instantaneous} hamiltonian, since it depends explicitly of time. 
Following Rhodes \cite{Rhodes-1989}, in order to diagonalize $\hat{H}(t)$ 
we introduce the time-dependent operator (we are using $\hbar = 1$)
\begin{equation}
\hat{b}(t)\equiv\sqrt{\frac{\omega(t)}{2}}\left(\hat{q}+i\frac{\hat{p}}{\omega(t)}\right) \, .
\label{eq:instantbasis}
\end{equation}
This operator and its hermitian adjoint satisfy the (equal-time) commutation relation:
\begin{equation}
\left[\hat{b}(t),\hat{b}^{\dagger}(t)\right]=1 \, .
\label{eq:InstComm}
\end{equation}
In terms of the above operators, the instantaneous hamiltonian, Eq.(\ref{eq:HOVF}), reads
\begin{equation}
\hat{H}(t)=\omega(t)\left[\hat{b}^{\dagger}(t)\hat{b}(t)+\frac{1}{2}\right] \, ,
\label{eq:HOVFbasis}
\end{equation}
therefore, it can be diagonalized by the eigenstates of the operator ${\hat b}^\dagger(t){\hat b}(t)$. 
Since operators $\hat{b}(t)$ and $\hat{b}^{\dagger}(t)$ serve 
to diagonalize the instantaneous hamiltonian, let us refer to them as the instantaneous representation. 

Let us now define the usual annihilation operator, namely
\begin{eqnarray}
\hat{a}&\equiv&\sqrt{\frac{\omega_{o}}{2}}\left(\hat{q}+i\frac{\hat{p}}{\omega_{o}}\right) \, , 
\label{eq:instantbasist0}
\end{eqnarray}
where $\omega_{o}\equiv\omega(t=0)$ and which satisfies the commutation relation $\left[\hat{a},\hat{a}^{\dagger}\right]=1$. 
Consequently, $\hat{a}^{\dagger}$ is the creation operator. 
Notice that $\hat{a}$ and $\hat{a}^{\dagger}$ coincide with $\hat{b}(t)$ and $\hat{b}^{\dagger}(t)$ at $t=0$, respectively, 
and they are used to diagonalize the hamiltonian at $t=0$, namely, $\hat{H}_{0}\equiv\hat{H}(t=0)$, as follows 
\begin{equation}
\hat{H}_{0} \left|n\right\rangle=\omega_{o}\left(\hat{n}+\frac{1}{2}\right)\left|n\right\rangle=\omega_{o}\left(n+\frac{1}{2}\right)\left|n\right\rangle \, ,
\label{eq:initialspec}
\end{equation}
with $\hat{n} =\hat{a}^{\dagger}\hat{a}$ the number operator, 
$\left|n\right\rangle=\frac{\left(\hat{a}^{\dagger}\right)^{n}}{\sqrt{n!}}\left|0\right\rangle$, and, 
consequently, $\left\{\left|n\right\rangle\right\}$ is the Fock space at $t=0$. 
Let us refer to $\hat{H}_{0}$ as the \textit{initial} hamiltonian, 
$\left\{\left|n\right\rangle\right\}$ as the initial basis and operators
$\hat{a}$ and $\hat{a}^{\dagger}$ as the initial representation.
Using Eq. (\ref{eq:instantbasist0}) to write position and momentum operators in the initial representation, 
and then replacing them in Eq. (\ref{eq:instantbasis}), we get
\begin{equation}
\hat{b}(t)=\cosh(\rho(t))\hat{a}+\sinh(\rho(t))\hat{a}^{\dagger} \, ,
\label{eq:Instannh}
\end{equation}
where
\begin{equation}
\rho(t)\equiv\frac{1}{2}\ln{\left(\frac{\omega(t)}{\omega_{o}}\right)} \, .
\label{eq:compdgr}
\end{equation}
From Eqs. (\ref{eq:InstComm}) and (\ref{eq:Instannh}),  we see that operators $\hat{b}(t)$ and $\hat{b}^{\dagger}(t)$ are 
given by real Bogoliubov transformations of operators $\hat{a}$ and $\hat{a}^{\dagger}$.
 
At this time it is convenient to introduce the so-called vacuum squeezed states 
(for an introduction on squeezed states we suggest Ref. \cite{Barnett-1997}). 
A single-mode vacuum squeezed state $\vert \zeta \rangle$ (referred to, henceforth, simply by squeezed state) 
of the initial hamiltonian $\hat{H}_{0}$, can be obtained by the application of the squeeze operator, ${\hat S}(\zeta)$, 
on the fundamental state of $\hat{H}_{0}$, 
namely, $\vert \zeta\rangle = {\hat S}(\zeta) \vert 0\rangle$, with
\begin{equation}
\label{gensqop}
\hat{S}(\zeta)\equiv\exp \left\{-\frac{\zeta}{2}\left.\hat{a}^{\dagger}\right.^{2}+\frac{\zeta^{*}}{2}\hat{a}^{2}\right\} \, ,
\end{equation}
where $\zeta = r e^{i\varphi}$ is a complex number and $*$ indicates complex conjugation. 
Notice that $\zeta$, and hence parameters $r$ and $\varphi$, determines uniquely the squeezed state. 
To interpret $r$ and $\varphi$, let us introduce 
the initial (since it is given in the initial representation) quadrature operator ${\hat Q}_\lambda$, defined by \cite{Barnett-1997}
\begin{equation}
\hat{Q}_{\lambda}=\frac{1}{\sqrt{2}}\left[e^{i\lambda} \hat{a}^{\dagger}+e^{-i\lambda} \hat{a}\right] \, ,
\label{eq:quadratureop}
\end{equation}
satisfying the commutation relation $[{\hat Q}_\lambda,{\hat Q}_{\lambda + \pi/2}] = i$. 
It is important to note that, rather than position and momentum operators, the initial quadrature operator depends on the initial frequency. 
In fact, from its definition and Eq. (\ref{eq:instantbasist0}), 
it is related to $\hat{q}$ and $\hat{p}$ by fixing the value of the quadrature, $\lambda$, as
$\hat{Q}_{\lambda=0} = (\hat{a}^{\dagger} +  \hat{a})/\sqrt{2} = \sqrt{\omega_{o}}\hat{q}\,$ and 
$\hat{Q}_{\lambda=\pi/2} = i(\hat{a}^{\dagger} -  \hat{a})/\sqrt{2}  =  \hat{p}/\sqrt{\omega_{o}}\,$. 
The HO is said to be squeezed if the variance of one quadrature is smaller than $\frac{1}{2}$, 
the so-called \textit{coherent limit}. 
It can be shown that the variance of the initial quadrature operator ${\hat Q}_\lambda$ 
in a squeezed state of the initial hamiltonian $\vert \zeta \rangle$, is given by \cite{Barnett-1997}
\begin{equation}
\left(\Delta Q_\lambda\right)^2 
=
\frac{e^{2r}}{2} \sin^{2}\left(\lambda-\varphi/2\right)+\frac{e^{-2r}}{2} \cos^{2}\left(\lambda-\varphi/2\right) \, . 
\label{eq:quadratureopvariance}
\end{equation}
Notice the explicit dependence of the variance with $r$ and $\varphi$. Even more, from the previous equation we see that 
\begin{equation}
 \frac{e^{-2r}}{2} \le \left(\Delta Q_\lambda\right)^2 \le  \frac{e^{2r}}{2}\, .
\label{eq:extvarval}
\end{equation}
From the above relation, parameter $r$ is called squeezing parameter (SP), since 
it determines how much the variance of the initial quadrature operator is below the coherent limit. 
Therefore, $r$ and $\left(\Delta Q_\lambda\right)^2$ are equivalent squeezing quantifiers. 
Parameter $\varphi$ is referred to as the squeezing phase (Sph). 

Coming back to the squeeze operator, it is unitary, and transforms the annihilation operator as \cite{Rhodes-1989, Barnett-1997}
\begin{equation}
\hat{S}(\zeta)\hat{a}\hat{S}(-\zeta)=\cosh(\left|\zeta\right|)\hat{a}+\frac{\zeta}{\left|\zeta\right|}\sinh(\left|\zeta\right|)\hat{a}^{\dagger} \, .
\label{eq:squeeandbass}
\end{equation}
Comparison of the last equation with Eq. (\ref{eq:Instannh}), allows us to write 
$\hat{b}=\hat{S}(\rho)\hat{a}\hat{S}(-\rho)$, where we have omitted the temporal 
dependence in the argument of the operators for simplicity of notation. 
We shall do that along the text whenever there is not risk of confusion. 
Notice that, from its definition in 
Eq. (\ref{eq:compdgr}), $\rho(t)$ is real valued. 
Nevertheless, we can identify the SP 
of $\hat{S}(\rho)$ with $\left|\rho(t)\right|$, and the Sph 
will be zero, for $\omega(t)\geq\omega_{o}$, and $\pi$, for $\omega(t)<\omega_{o}$.
Using the above results, Eq. (\ref{eq:HOVFbasis}) can be written as:
\begin{equation}
\hat{H}(t)=\frac{\omega(t)}{\omega_{o}}\hat{S}\left(\rho(t)\right)\hat{H}_{0}\hat{S}\left(-\rho(t)\right) \, .
\label{eq:HOVFnewold}
\end{equation}
Accordingly, we can say that the unitary transformation carried out by $\hat{S}\left(\rho\right)$, 
up to the {time-dependent} factor $\omega(t)/\omega_{o}$, 
has the effect to \textit{update} the initial hamiltonian to the instantaneous one. 
From the above result and Eq. (\ref{eq:initialspec}), it follows that
\begin{equation}
\hat{H}(t) \left(\hat{S}(\rho(t))\left|n\right\rangle\right)=\omega(t)\left(n+\frac{1}{2}\right)\left(\hat{S}(\rho(t))\left|n\right\rangle\right).
\label{eq:insteigen}
\end{equation}
Hence, $\hat{S}(\rho(t))\left|n\right\rangle$ is an eigenstate of $\hat{H}(t)$. 
Let us write it as 
\begin{equation}
\left|n\right\rangle_{t}\equiv\hat{S}(\rho(t))\left|n\right\rangle .
\label{eq:instsqueezedstate}
\end{equation}
A direct consequence is that the fundamental state in the instantaneous basis is
a squeezed state in the initial basis, explicitly, 
\begin{equation}
\left|0\right\rangle_{t}=\sqrt{\mbox{sech}(\rho)}\sum_{n=0}^{\infty}\frac{\sqrt{(2n)!}}{n!}
\left[-\frac{1}{2}\tanh(\rho)\right]^{n}\left|2n\right\rangle \, .
\label{eq:squeezedsuperp}
\end{equation}
The above result will be useful in the next section, when discussing the behavior 
of the SP in a frequency transition. 
If we define $\hat{n}(t) =\hat{b}^{\dagger}(t)\hat{b}(t)$ as the instantaneous number operator, then 
$\left|n\right\rangle_{t}=\frac{\left(\hat{b}^{\dagger}(t)\right)^{n}}{\sqrt{n!}}\left|0\right\rangle_{t}$, and  
$\left\{ \left|n\right\rangle_{t} \right\}$ is the Fock space at time $t$.  
Henceforth, $\left\{ \left|n\right\rangle_{t} \right\}$ will be refereed to as the instantaneous basis
\footnote{This basis is also known as adiabatic basis \cite{Del-Campo-2011,guery-2019}.}. 
Notice that the initial and instantaneous bases coincide at $t=0$ and whenever $\omega(t)=\omega_{o}$. 

At this point, it is convenient to introduce the generators of the \textit{su}(1,1) Lie algebra in the following form:
\begin{equation}
\hat{K}_{+} := \frac{\hat{a}^{\dagger^{2}}}{2}, \:\: \hat{K}_{-} := \frac{\hat{a}^{2}}{2} \:\:\:\: \mbox{and} \:\:\:\: \hat{K}_{c} := 
\frac{\hat{a}^{\dagger}\hat{a}+\hat{a}\hat{a}^{\dagger}}{4}\, ,
\label{eq:LieAlgebraGen}
\end{equation}
satisfying the commutation relations
\begin{equation}
\left[\hat{K}_{+},\hat{K}_{-}\right]=-2\hat{K}_{c} \:\: \mbox{and} \:\: \left[\hat{K}_{c},\hat{K}_{\pm}\right]=\pm\hat{K}_{\pm}\, .
\label{eq:algebraK}
\end{equation}
Using Eqs. (\ref{eq:Instannh}) and (\ref{eq:LieAlgebraGen}), the instantaneous hamiltonian in Eq. (\ref{eq:HOVFbasis}) can be written in the 
initial representation as
\begin{eqnarray}
\hat{H}(t)&=&2\omega(t)\cosh(2\rho(t))\hat{K}_{c} + \nonumber \\
&& + \omega(t)\sinh(2\rho(t))\left(\hat{K}_{+}+\hat{K}_{-}\right) \, .
\label{eq:HamLie}
\end{eqnarray}
Notice that in the above form of the hamiltonian, all the time-dependence is encoded in the coefficients and not in the operators, 
as it occurs when the instantaneous representation is used (see Eq. (\ref{eq:HOVFbasis})).

In the following discussion we shall use the solution for a TDHO developed in Ref. \cite{DMTibaTDHO-2020}, 
for an arbitrary frequency function and any initial state, 
but assuming that the HO is initially in its fundamental state. 
This solution is given in the initial basis, 
and is extremely convenient to do numerical calculations. 
In this reference the authors considered time splitting in $N$ intervals of 
equally small enough size $\tau=t/N$, such 
that the time-dependent frequency $\omega(t)$ 
and, consequently, the hamiltonian Eq. (\ref{eq:HamLie}), 
can be considered constant in each interval. 
Using algebraic methods the authors showed that
the time evolution operator (TEO) of the system
can be written as 
a squeeze operator in the initial representation \textit{i.e.}, $\hat{U}(t,0)={\hat S}(z(t))$.
Consequently, the state of the system at time $t$ is a squeezed state of the initial hamiltonian
$\vert z(t)\rangle = \hat{U}(t,0) \vert 0\rangle$, with $z(t) = r(t) e^{i\varphi (t)}$. 
Defining the value of the frequency in the $j$-th time-interval as $\omega_{j} :=\omega(j\tau)$, with $j=1,2,...,N$, 
the complex variable $z(t)$ that characterizes the state can be calculated from 
the following recurrence relation \cite{DMTibaTDHO-2020}:
\begin{equation}
\chi_{N}=a_{N}+\frac{\chi_{N-1}b_{N}}{1-\chi_{N-1}a_{N}}
\label{eq:reccurence1}
\end{equation}
where 
\begin{eqnarray}
\label{truej9}
a_{j}&=&\frac{-i \sinh(2\rho_{j})\sin(\omega_{j}\tau)}{\cos(\omega_{j}\tau)+i\cosh(2\rho_{j})\sin(\omega_{j}\tau)} \, ,\\[3pt]
\label{truej8}
b_{j}&=&\left(\cos(\omega_{j}\tau)+i\cosh(2\rho_{j})\sin(\omega_{j}\tau)\right)^{-2}
\end{eqnarray}
and $\rho_{j}=\frac{1}{2}\ln{\left(\frac{\omega_{j}}{\omega_{o}}\right)}$. 
Expressing $\chi$ in its polar form, $\chi=\left|\chi \right| e^{i \theta}$, 
the state vector of the system $\vert z(t)\rangle = \vert r(t) e^{i\varphi (t)}\rangle$ 
is obtained by the following identifications
\begin{equation}
r(t)=\tanh^{-1}\left|\chi_{N}\right| \:\:\:\: \mbox{and}\:\:\:\: \varphi(t)=\theta_{N}\pm\pi \, ,
\label{eq:phaseparamsque}
\end{equation}
with $N$ big enough to guarantee convergence in the values of the SP, $r(t)$, and the Sph, $\varphi(t)$. 

Recall that Eq.(\ref{eq:phaseparamsque}) corresponds to the solution using the eigenstates of $\hat{H}_{0}$. 
We shall continue our discussion by describing the system using the eigenstates of $\hat{H}(t)$. 
To do that, we shall write the state of the system, $\vert z(t)\rangle$, in the instantaneous basis.
Using the unitary property of the squeeze operator and Eq. (\ref{eq:instsqueezedstate}), 
the state of the system can be written as
\begin{eqnarray}
\label{eq:instbasfin}
\vert z(t)\rangle&=& \hat{S}\left(z(t)\right) \vert 0\rangle \cr\cr
								 &=& \hat{S}\left(z(t)\right) \hat{S}\left(-\rho(t)\right) \left|0 \right\rangle_{t} \, .
\end{eqnarray}
The solution in this approach depends on rewriting the product of the squeeze operators, $\hat{S}\left(z(t)\right) \hat{S}\left(-\rho(t)\right)$, 
in the instantaneous representation. To do this, 
first note that the Bogoliubov transformation given by Eq. (\ref{eq:Instannh}) (and its hermitian adjoint) can be inverted to obtain
\begin{equation}
\hat{a}=\left.\Gamma_{1}\right.\hat{b}-\left.\Gamma_{2}\right.\hat{b}^{\dagger} \:\:\: \mbox{and} \:\:\: 
\hat{a}^{\dagger}=\left.\Gamma_{1}\right.\hat{b}^{\dagger}-\left.\Gamma_{2}\right.\hat{b} \, ,
\label{eq:Instannhinv}
\end{equation}
where we defined
\begin{equation}
\Gamma_{1}=\cosh(\rho(t)) \:\:\:\: \mbox{and}\:\:\:\: \Gamma_{2}=\sinh(\rho(t)) \, .
\label{eq:Bogcoeff}
\end{equation}
Also, note that using the commutation relation, Eq. (\ref{eq:InstComm}), we can define the generators of the \textit{su}(1,1) Lie algebra as:
\begin{equation}
\hat{T}_{+} := \frac{\hat{b}^{\dagger^{2}}}{2}, \:\: \hat{T}_{-} := \frac{\hat{b}^{2}}{2} \:\:\:\: \mbox{and} 
\:\:\:\: \hat{T}_{c} := \frac{\hat{b}^{\dagger}\hat{b}+\hat{b}\hat{b}^{\dagger}}{4}\, ,
\label{eq:LieAlgebraGeninsbas}
\end{equation}
satisfying commutation relations analogous to those of Eq. (\ref{eq:algebraK}), just replacing $\hat{K}\rightarrow\hat{T}$. 
Now, using Eqs. (\ref{eq:Instannhinv}) and (\ref{eq:LieAlgebraGeninsbas}) in Eq. (\ref{gensqop}), 
a generic squeeze operator in the initial representation is transformed to the instantaneous representation as
\begin{equation}
\hat{S}\left(\zeta\right)=\exp\left\{\lambda_{+}\hat{T}_{+}+\lambda_{c}\hat{T}_{c}+\lambda_{-}\hat{T}_{-}\right\} \, ,
\label{eq:finalinstbasfin}
\end{equation}
where we defined
\begin{eqnarray}
\lambda_{+}&=&\left(\zeta^{*}\left.\Gamma_{2}\right.^{2}-\zeta\left.\Gamma_{1}\right.^{2}\right)=-\lambda^{*}_{-} \, , \cr\cr
\lambda_{c}&=&2 \Gamma_{1}\Gamma_{2} \left(\zeta-\zeta^{*}\right)\, .
\label{eq:lambdasalgebraT}
\end{eqnarray}
Consequently, operators $\hat{S}\left(z(t)\right)$ and $\hat{S}\left(-\rho(t)\right)$ in Eq. (\ref{eq:instbasfin}) 
are transformed to the instantaneous representation by substituting in Eq. (\ref{eq:finalinstbasfin})
$\zeta\equiv z(t)$ and $\zeta\equiv \rho(t)$, respectively. 
The presence of the term $\lambda_{c}\hat{T}_{c}$ in Eq. (\ref{eq:finalinstbasfin}) 
implies that a generic squeeze operator in the initial representation is not, in general, a 
squeeze operator in the instantaneous one. 
However, since $\rho=\rho^{*}$, operator $\hat{S}\left(-\rho(t)\right)$ in the instantaneous representation 
is given by
\begin{equation}
\hat{S}\left(-\rho\right)=\exp\left\{\rho\hat{T}_{+}-\rho\hat{T}_{-}\right\} \, .
\label{eq:finalrhoinstbasfin}
\end{equation}
which turns out to be a squeeze operator. 
Now, since $\hat{S}\left(z(t)\right)$ and $\hat{S}\left(-\rho(t)\right)$ can be 
identified as elements of the \textit{su}($1,1$) Lie algebra, 
therefore we can use new BCH-like relations \cite{DMTibaBCH-2020} 
to calculate their product as 
\begin{equation}
\hat{S}\left(z\right)\hat{S}\left(-\rho\right)=e^{\alpha\hat{T}_{+}}e^{\ln(\beta)\hat{T}_{c}}e^{\gamma\hat{T}_{-}} \, ,
\label{eq:compoBCHfin}
\end{equation}
where
\begin{eqnarray}
\label{beta2}
&&\alpha=\Lambda_{+}+\frac{\Gamma_{2}\Lambda_{c}}{\Gamma_{1}-\Gamma_{2}\Lambda_{-}} \, , \:\:\:\:\:\:
\beta=\frac{\Lambda_{c}}{\left(\Gamma_{1}-\Gamma_{2}\Lambda_{-}\right)^{2}} \,   \nonumber\\
&&\:\:\:\:\:\:\:\:\:\:\:\:\:\:\:\:\:\:\:\:\:\mbox{and} \:\:\: \gamma=\frac{\Gamma_{1}\Lambda_{-}-\Gamma_{2}}{\Gamma_{1}-\Gamma_{2}\Lambda_{-}} \, ,
\end{eqnarray}
with
\begin{eqnarray}
\label{truej4}
\Lambda_{+}&=&\frac{\left( e^{-i\varphi}\left.\Gamma_{2}\right.^{2} - e^{i\varphi}\left.\Gamma_{1}\right.^{2} \right)\sinh(r)}
{\cosh(r) - \Gamma_{1}\Gamma_{2}\left( e^{i\varphi} - e^{-i\varphi}\right)\sinh(r)} \,  , \nonumber\\[5pt]
\Lambda_{-}&=&\frac{\left( e^{-i\varphi}\left.\Gamma_{1}\right.^{2} - e^{i\varphi}\left.\Gamma_{2}\right.^{2} \right)\sinh(r)}
{\cosh(r) - \Gamma_{1}\Gamma_{2}\left( e^{i\varphi} - e^{-i\varphi}\right)\sinh(r)} \,  , \nonumber\\[5pt]
\Lambda_{c}&=&\left(\cosh(r)-\Gamma_{1}\Gamma_{2}\left( e^{i\varphi} - e^{-i\varphi}\right) \sinh(r)\right)^{-2}.
\end{eqnarray}
Finally, using the following results
\begin{eqnarray}
\hat{T}_{-}\left|n\right\rangle_{t}=\frac{1}{2}\sqrt{n(n-1)}\left|n-2\right\rangle_{t} \, , \cr\cr
\hat{T}_{+}\left|n\right\rangle_{t}=\frac{1}{2}\sqrt{(n+1)(n+2)}\left|n+2\right\rangle_{t} \, , \cr\cr
\hat{T}_{c}\left|n\right\rangle_{t}=\frac{1}{2}\left(n+\frac{1}{2}\right)\left|n\right\rangle_{t} \, ,
\label{eq:rules}
\end{eqnarray}
and the well known expansion $e^{\hat{A}}=\sum_{n=0}^{\infty}{\frac{1}{n!}\hat{A}^{n}}$ 
(valid for a general operator $\hat{A}$), substitution of Eq. (\ref{eq:compoBCHfin}) in Eq. (\ref{eq:instbasfin}) yields
\begin{equation}
\vert z(t)\rangle=\sqrt{\left|\beta \right|^{1/2}}\sum_{n=0}^{\infty}\frac{\sqrt{(2n)!}}{n!}
\left[\frac{1}{2}\left|\alpha\right|e^{i\vartheta}\right]^{n}\left|2n\right\rangle_{t},
\label{eq:wfinalg2}
\end{equation}
where the overall phase was removed by the redefinitions
\begin{equation}
\alpha=\left|\alpha\right|e^{i\vartheta} \:\:\:\: \mbox{and}\:\:\:\: \beta=\left|\beta\right|e^{i\upsilon}.
\label{eq:phasecoeffg2}
\end{equation}
It can be 
checked that the following relation is satisfied
\begin{equation}
\left|\alpha\right|^{2}+\left|\beta\right| = 1 \, ,
\label{eq:modulustrig}
\end{equation}
and, accordingly, the state vector of the system in the instantaneous basis can also be identified as a squeezed state, 
but now of the instantaneous hamiltonian $\hat{H}(t)$. 
Let us call the state vector on this basis as $\left| \xi(t) \right\rangle_{t}$, with $\xi(t)=R(t) e^{i\Phi(t)}$. 
Therefore, $R(t)$ and $\Phi(t)$ are the corresponding SP and Sph of $\left| \xi(t) \right\rangle_{t}$ and they can 
be calculated from the complex variable $\alpha$ in Eq. (\ref{beta2}) as 
\begin{equation}
R(t)=\tanh^{-1}\left|\alpha\right| \:\:\:\: \mbox{and}\:\:\:\: \Phi(t)=\vartheta\pm\pi \, .
\label{eq:PHAseparamsqueBas}
\end{equation}
In order to interpret $R(t)$ and $\Phi(t)$, let us introduce 
the instantaneous quadrature operator, $\hat{Q'}_{\lambda}$, given by
\begin{equation}
\hat{Q'}_{\lambda}(t)=\frac{1}{\sqrt{2}}\left[e^{i\lambda} \hat{b}^{\dagger}(t)+e^{-i\lambda} \hat{b}(t)\right] \, ,
\label{eq:quadratureopact}
\end{equation}
satisfying the commutation relation $[{\hat Q'}_\lambda,{\hat Q'}_{\lambda + \pi/2}] = i$. 
The above operator is obtained by making the changes $\hat{a}\rightarrow \hat{b}$ and 
$\hat{a}^{\dagger}\rightarrow \hat{b}^{\dagger}$ in Eq. (\ref{eq:quadratureop}). 
Further, the instantaneous and initial quadrature operators are related by an unitary 
transformation carried out by $\hat{S}\left(\rho\right)$, namely, $\hat{Q'}_{\lambda}(t)=\hat{S}(\rho(t))\hat{Q}_{\lambda}\hat{S}(-\rho(t))$.
Since $\hat{Q'}_{\lambda}$ is expressed in the instantaneous representation, 
it can be shown that its variance is analogous to Eq. (\ref{eq:quadratureopvariance}), 
but this time with the SP and Sph given by Eqs. (\ref{eq:PHAseparamsqueBas}), namely,
\begin{equation}
\left(\Delta Q'_{\lambda}\right)^2 =
\frac{e^{2R}}{2} \sin^{2}\left(\lambda-\Phi/2\right)+\frac{e^{-2R}}{2} \cos^{2}\left(\lambda-\Phi/2\right) \, . 
\label{eq:quadratureopvarianceinst}
\end{equation}
Moreover, from the previous equation we see that 
\begin{equation}
 \frac{e^{-2R}}{2} \le \left(\Delta Q'_\lambda\right)^2 \le  \frac{e^{2R}}{2}\, ,
\label{eq:extvarvalinst}
\end{equation}
which provides an equivalent relation between $R(t)$ and $\left(\Delta Q'_{\lambda}\right)^2$ 
as that written in Eq. (\ref{eq:extvarval}) for $r(t)$ and $\left(\Delta Q_{\lambda}\right)^2$. 
Recall that the variance of the instantaneous quadrature operator could also be calculated using 
any other basis, including the initial one. 

We have shown that the state vector of a HO, initially in its fundamental state and whose frequency begins to change in time, 
evolves to a squeezed state of the initial hamiltonian, when the initial basis is used to describe the system. 
Even more, when employed the instantaneous basis to describe the system, the state vector of the HO also evolves 
to a squeezed state, but of the instantaneous hamiltonian. 
In the following section we shall study the relation 
between squeezing and adiabaticity in a frequency transition by 
varying the frequency rate. 




\section{Squeezing and adiabaticity in a frequency transition}\label{SvsA}

As we have shown in the previous section, the state vector of the HO with a time-dependent frequency 
can be written in two different, but equivalent ways, namely: as a squeezed state relative to 
the initial hamiltonian, $\vert r(t) e^{i\phi(t)}\rangle$, or as a squeezed state relative to 
the instantaneous hamiltonian, $\vert R(t) e^{i\Phi(t)}\rangle$.
Here, we shall calculate the SP's, $r(t)$ and $R(t)$, 
for a given class of frequency transitions, which are monotonic functions 
connecting the initial and final frequencies of the HO. 
The contrast between these two bases descriptions will allow us 
to discuss some subtleties and common inaccuracies present in the literature 
related to the interpretation of the adiabatic theorem. 
We recall that the numerical implementation of our previous results, 
obtained in section II, enables us to determine the state vector of the system 
as accurate as desired for any frequency function, which shows the robustness of our method.

We model the frequency transition with the following family of hyperbolic tangent functions \cite{FUJII-2011}: 
\begin{equation}
\omega(t) = \frac{\omega_{f}+\omega_{o}}{2}+\frac{\left(\omega_{f}-\omega_{o}\right)}{2}
\tanh\left(\frac{t-t_{0}}{\epsilon}\right)\, ,
\label{eq:frequency1}
\end{equation}
where $\omega_o$ and $\omega_f$ are the HO frequencies achieved asymptotically in the limits of remote past 
($t\rightarrow - \infty$)  and  distant future ($t\rightarrow + \infty$), respectively,  
$t_0$ is the instant for which the HO frequency assumes the halfway value between 
$\omega_o$ and $\omega_f$, and $\epsilon$ is, by assumption, a non-negative continuous parameter, 
with dimensions of time, that can be adjusted appropriately for our purposes. 
This kind of time-dependence on the frequency makes it possible 
to avoid possible numerical divergences \cite{FUJII-2011}, 
while the values of $\epsilon$ 
control the transition rate. 
Indeed, from Eq. (\ref{eq:frequency1}) 
%
\begin{equation}
\epsilon = \frac{\omega_{f}-\omega_{o}}{2\dot{\omega}(t_0)} \, ,
\label{eq:epsiveloci}
\end{equation}
where the overdot indicates time derivative. 
Accordingly, the limiting case $\epsilon\rightarrow 0$ corresponds to a sudden change at $t = t_0$, 
and the case $\epsilon \rightarrow \infty$ corresponds to a totally smooth change. 
Parameter $\epsilon$ also allows us to define the \textit{transition interval}, denoted by $\mathcal{I}$, 
as the time interval (around $t_0$) in which the frequency changes appreciably. 
Due to the features of the hyperbolic function and the set of parameters we shall use in the following sections, 
we choose $\mathcal{I}\equiv\left(t_0 - 3\epsilon, t_0 + 3\epsilon \right)$. 
Without any loss of generality, in this section 
we set for numerical calculations $\omega_{o}=1$, $\omega_{f}=3$ 
and $t_{0}=10$ (in arbitrary units). 
Moreover, although the HO frequency coincides with $\omega_o$ only in the remote past, 
we will consider that, for $t<t_0 - 3\epsilon$, 
$\omega(t)$ can be safely approximated by $\omega_o$. 
Accordingly, the state vector of the HO can be considered 
as being its fundamental state for $t< t_0 -3\epsilon$. 
Similarly, we assume that $\omega(t)$  has already  achieved its asymptotic value $\omega_f$ 
for $t>t_{0}+3\epsilon$.
In  panel (a) of Fig. \ref{fig:FreqVsTime} we plot the frequency given by Eq. (\ref{eq:frequency1}) 
as a function of time for different values of $\epsilon$, namely, $\epsilon = 10^{-3}, 0.5, 1.0$ and $1.5$. 
Clearly, from this graphic $\epsilon=10^{-3}\sim 0.0$ corresponds to the fastest transition while 
$\epsilon=1.5$ corresponds to the slowest one. 
As we shall see, the previous choices for $t_0$ and $\epsilon$ will be quite good for our discussion. 


\subsection{SP in the initial basis} \label{SPinint}

\begin{figure}
\centering
\includegraphics*[width=8.0cm]{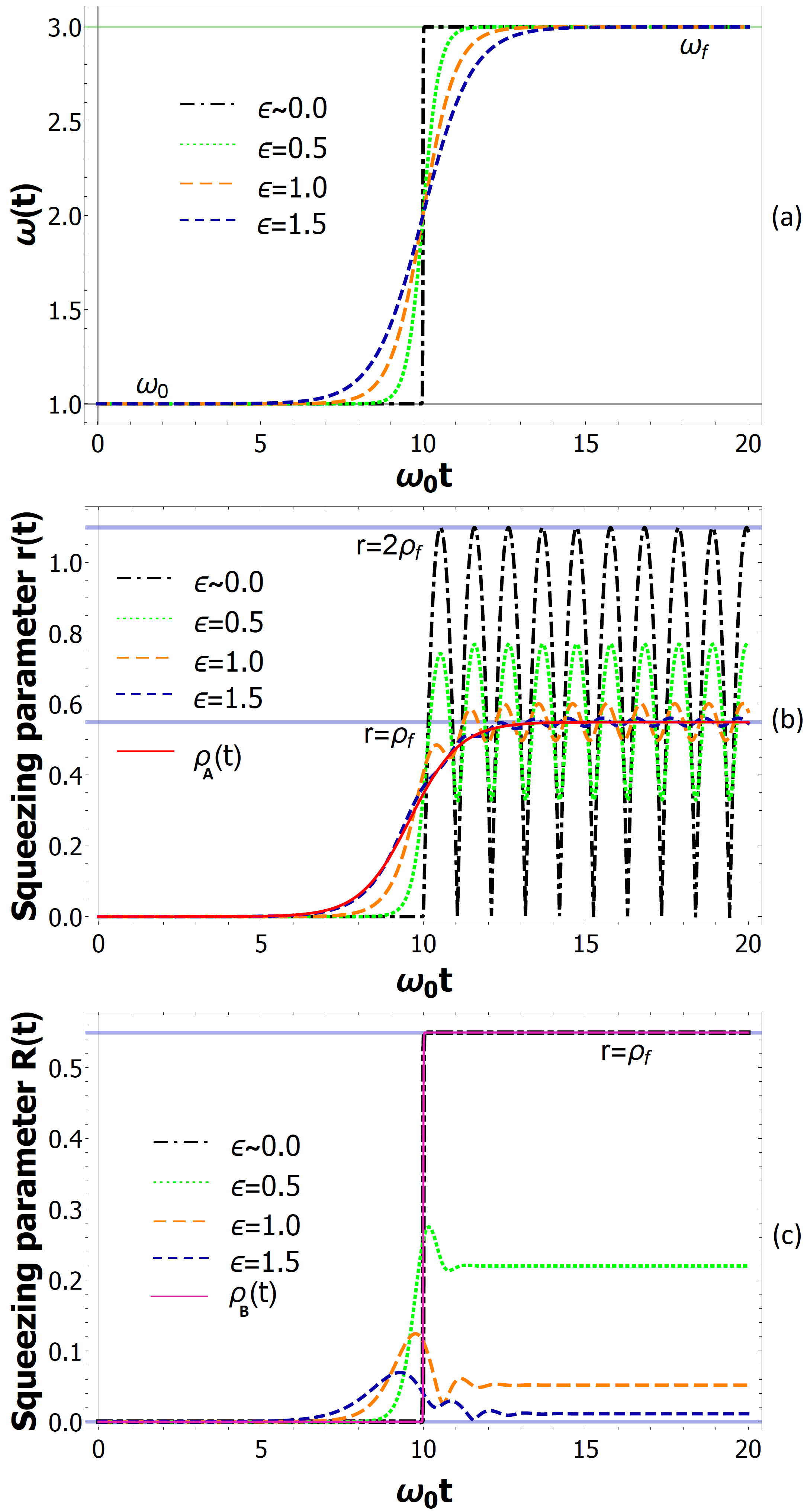}
\caption{(a) HO frequency  given by Eq. (\ref{eq:frequency1}) as a function of time;
(b) time evolution of the SP in the initial basis and (c) time evolution of the SP in the instantaneous basis. 
In all panels, different line patterns mean different values of $\epsilon$; 
$\epsilon \sim 0.0$ (dot-dashed line and sudden limit), $\epsilon = 0.5$ (dotted line), 
$\epsilon =1.0$ (long dashed line) and $\epsilon= 1.5$ (short dashed line and adiabatic limit). 
The solid lines $\rho_{A}(t)$  (in panel (b)) and $\rho_{B}(t)$ (in panel (c)) are the plots of $\rho(t)$ 
(Eq. (\ref{eq:compdgr})) as a  function of $\omega_o t$,  with  $\omega(t)$ being the time-dependent frequency 
considered for $\epsilon = 1.5$ and $\epsilon \sim 0.0$, respectively.}
\label{fig:FreqVsTime}
\vskip -0.5cm
\end{figure}

Using Eqs. (\ref{eq:reccurence1})-(\ref{eq:phaseparamsque}), we calculate 
the SP of the HO at a generic instant $t>0$ in the initial basis, 
namely, $r(t)$, for the time-dependent frequency given in Eq. (\ref{eq:frequency1}). 
We plot $r(t)$ as a function of time in Fig. \ref{fig:FreqVsTime}(b), 
employing the same values for $\epsilon$ as those used in Fig. \ref{fig:FreqVsTime}(a). 
From this figure it can be noted that, once the transition has occurred ($t> t_0 + 3\epsilon$), 
$r(t)$ acquires a periodic behavior. 
The faster the transition is (the smaller $\epsilon$ is), the earlier the periodic behavior is achieved. 
Also, for $t> t_0 + 3\epsilon$ both, the period and the halfway point between the minimum and 
maximum values of $r(t)$ (referred to as midpoint of $r(t)$ from now on), 
are the same for all transitions. With this fact in mind, and recalling that the  
case of a jump in the frequency ($\epsilon=0$) has analytical solution, 
let us infer some relevant characteristics of 
a general transition (arbitrary $\epsilon$), from this particular limiting case. 
For a frequency jump, the SP can be written as \cite{DMTibaBJP-2020}:
\begin{equation}
r_{\epsilon=0}(t) = \mbox{arcosh}{\sqrt{1+\left(\frac{\left.\omega_{f}\right.^{2}-\left.\omega_{o}\right.^{2}}
{2\omega_{o}\omega_{f}}\right)^2 \sin^{2}\left(\omega_{f}t \right)}} \, .
\label{eq:SPfreqjump}
\end{equation}
A direct inspection of the above equation shows that the period of $r_{\epsilon=0}(t)$ is 
$T=\pi/\omega_f$, since $r_{\epsilon=0}(t + \pi/\omega_f) = r_{\epsilon=0}(t)$. Therefore, 
the period of $r(t)$ for any 
finite $\epsilon$ is also $T$. Now, from Eq. (\ref{eq:SPfreqjump}) it 
can be shown that the (periodic) maximum of $r_{\epsilon=0}(t)$ is $2\rho_{f}$ (upper horizontal line) 
with $\rho_{f}=\frac{1}{2}\ln{\left(\frac{\omega_{f}}{\omega_{o}}\right)}=\rho(t> t_0 + 3\epsilon)$ 
(see Eq. (\ref{eq:compdgr})), while the (periodic) minimum of $r_{\epsilon=0}(t)$ is zero. Hence, 
once the transition has occurred, the midpoint of $r(t)$ is $\rho_{f}$ (lower horizontal line) 
for any 
finite $\epsilon$.
It can also be noted from Fig. \ref{fig:FreqVsTime}(b) that, the slower the transition, the smaller 
the amplitude of $r(t)$ and the smoother the oscillations.
In summary, the period and the midpoint of the $r(t)$ oscillations 
depend only on the initial and final frequency values, 
while the amplitude of oscillations is a signature of the transition rate. 
%

The behavior of the amplitude can be understood under the light of the adiabatic theorem. 
Essentially, this theorem states that, along an adiabatic transition, the instantaneous state 
of the system remains the same \cite{Born-Fock-1928}. 
Recall that the instantaneous basis is the set of eigenstates of $\hat{H}(t)$ 
and the dynamics begins with 
the system in its fundamental state. Accordingly, 
the state of the system in the instantaneous basis should approximate the fundamental 
of $\hat{H}(t)$ as the transition becomes slower.
Also, recall that the fundamental state of $\hat{H}(t)$ in the instantaneous basis 
is a squeezed state in the initial basis, namely, 
$\left|0\right\rangle_{t}=\hat{S}(\rho(t))\left|0\right\rangle=\left|\rho(t)\right\rangle$ (see Eq. (\ref{eq:squeezedsuperp})). 
Hence, when the system approaches the adiabatic limit, the state vector in the initial basis must satisfy 
$\left|r(t) e^{i\varphi(t)}\right\rangle\rightarrow\left|\rho(t)\right\rangle$, 
or equivalently, $r(t)\rightarrow \left|\rho(t)\right|$ and $\varphi(t)\rightarrow 2n\pi$, with $n=0,1,2,3,...$. 
Furthermore, since $\rho(t)$ can take only positive values because we choose $\omega(t)>\omega_{o}$, 
then we should have $r(t)\rightarrow\rho(t)$. 
And this is exactly what occurs: 
as we increase the value of $\epsilon$, $r(t)$ 
approaches $\rho(t)$. Moreover, as it can be seen from Fig. \ref{fig:FreqVsTime}(b), for $\epsilon=1.5$ 
(short dashed line) $r(t)$ almost coincides with $\rho_{A}(t)$ (solid line), 
the latter being $\rho(t)$ with $\omega(t)$ given by Eq. (\ref{eq:frequency1}) 
considered with $\epsilon = 1.5$. 
Although this description is consistent with the adiabatic theorem, it may seem a little bit counterintuitive 
the fact that the SP has a non zero value in an adiabatic transition. However, this is a direct consequence 
of the basis chosen for describing the state of the system and, as we shall see in the following subsection, 
when employed the instantaneous basis, the SP is zero after an adiabatic transition.

\subsection{SP in the instantaneous basis} \label{SPinst}

To get a direct and more intuitive interpretation of the adiabatic theorem, 
we shall now describe the dynamics of the SP in the instantaneous basis, namely, $R(t)$. 
Using Eqs. (\ref{eq:Bogcoeff}), (\ref{beta2}), (\ref{truej4}) and (\ref{eq:PHAseparamsqueBas}), 
we calculate $R(t)$ for the frequency given by Eq. (\ref{eq:frequency1}). Its behavior as 
a function of time is presented in Fig. \ref{fig:FreqVsTime}(c), with the same values for $\epsilon$ as before. 
As it can be seen in this figure, in contrast to what happens with $r(t)$, 
after the transition has occurred $R(t)$ has a constant behavior, instead of a periodic one. 
Notice that the maximum value reached by $R(t)$ is $\rho_{f}$, 
occurring for a jump in the frequency, which is the value of $r(t)$ for the adiabatic limit.
Naively, one could think that, for a frequency jump, the state vector of the 
system suffers a discontinuity at $t=t_{0}$. 
However, this is not the case, since, as Janszky points out in Ref. \cite{JANSZKY-1994}, 
even in a frequency jump the TEO satisfy $\lim_{t\rightarrow t_0}\hat{U}_{\epsilon\rightarrow0}(t,t_{0})=1$. 
It is worth emphasizing that, although the periodic behavior of the squeezing 
seems to have been lost, it will actually be present trough the squeezing phase. 
Indeed, it can be shown that, 
once the transition has occurred, 
the variance of $\hat{Q'}_{\lambda}(t)$ for $\lambda=0$ and $\lambda=\pi/2$, 
will oscillate around the coherent limit ($1/2$) with period $T$.

Another interesting fact in this description is that $R(t)$ approaches $\rho(t)$ in the sudden case, 
and not in the adiabatic one as occur for $r(t)$. 
In fact, as it can be noted in Fig. \ref{fig:FreqVsTime}(c), for $\epsilon\sim 0$ 
(dot-dashed line and sudden limit), $R(t)$ and $\rho_{B}(t)$ are almost superposed, 
the latter being $\rho(t)$ with $\omega(t)$ considered with the lowest value of epsilon. 
Finally, on this basis the adiabatic theorem indicates that $R(t)$ should approach zero as the 
transition slows down, since $\left| \xi(t) \right\rangle_{t}\rightarrow \left| 0 \right\rangle_{t}$ 
as $\epsilon \rightarrow \infty$. This behavior is evidenced in Fig. \ref{fig:FreqVsTime}(c) and, 
therefore, this basis affords a direct and intuitive interpretation of the adiabatic theorem. 

Notice that, regardless of the basis chosen to represent the state, the behavior of the 
SP provides a way to interpret the adiabatic theorem for our system. 
In fact, as we have shown, and as it is well known \cite{Husimi-1953}, 
the natural basis for describing this theorem is the instantaneous one. 
However, in case one chooses another basis, all one has to do 
is to interpret the adiabatic theorem accordingly.  
In the following subsection we discuss some common inaccuracies in the interpretation of the adiabatic theorem 
for this system, and how they are related to the choice of the basis.


\subsection{Discussion on squeezing and the adiabatic theorem}

In section II, we have shown how the squeezing parameters and the variance 
of the quadrature operators are related (see Eqs. (\ref{eq:extvarval}) and (\ref{eq:extvarvalinst})). Indeed, $r(t)$ and 
$\left(\Delta Q_{\lambda}\right)^{2}$ are equivalent quantifiers of the squeezing in the initial basis, 
while $R(t)$ and $\left(\Delta Q'_{\lambda}\right)^{2}$ fulfill the same role in the instantaneous basis. 
Nevertheless, in the literature all such quantifiers have been used to describe the squeezing in the TDHO, and, 
in some cases, there has been common inaccuracies in the interpretation of the adiabatic theorem. 
Here we want to address such inaccuracies, indicate their possible source, and show how they can be solved. 

Let us initially recall that position and momentum operators are, up to a constant, 
proportional to the initial quadrature operator, as we showed in Section \ref{IB}. 
Accordingly, $r(t)$ describes the squeezing associated to the variance of all these operators. 
Also, recall that in this initial basis perspective squeezing will inevitably be generated 
for any rate of frequency change, sudden or smooth, since $r(t)\neq 0$. 
Indeed, as we showed in Section \ref{SPinint}, 
for an adiabatic the transition 
$r(t)\rightarrow \left|\rho(t)\right|$. 
This non-vanishing squeezing is a consequence of being out of the instantaneous basis, 
and represents a kind of relative squeezing. 
On the other hand, in the instantaneous basis adiabatic changes in the frequency do not produce squeezing, 
and consequently this basis seems very 
convenient to work with. 
Nonetheless, some authors, that have also analysed the TDHO in the adiabatic limit and within 
the initial basis perspective, incorrectly have conclude that adiabatic changes 
in the frequency do not produce this `relative' squeezing (see for instance \cite{Rhodes-1989, Kumar-1991, Averbukh-1994}). 
We believe this inaccuracy has its source in the interpretation of $r(t)$ as $R(t)$, or equivalently, 
of $\left(\Delta Q_{\lambda}\right)^{2}$ as $\left(\Delta Q'_{\lambda}\right)^{2}$. 

In order to clarify this issue, let us indicate how $\left(\Delta Q_{\lambda}\right)^{2}$ and 
$\left(\Delta Q'_{\lambda}\right)^{2}$ are related to each other for the relevant quadratures. 
First, notice that position and momentum operators can also be written as functions of the 
instantaneous quadrature operator. Explicitly, using Eqs. (\ref{eq:instantbasis}) and 
(\ref{eq:quadratureopact}) it can be shown that 
$\hat{Q'}_{\lambda=0}(t) = \sqrt{\omega(t)}\hat{q}\,$ and 
$\hat{Q'}_{\lambda=\pi/2}(t) = \hat{p}/\sqrt{\omega(t)}$. Using the similar expressions that we found in 
Section II for position and momentum operators in terms of the initial quadrature operator, we can write 
$\hat{Q'}_{\lambda=0}(t) = \sqrt{\frac{\omega(t)}{\omega_o}}\hat{Q}_{\lambda=0}\,$ and 
$\hat{Q'}_{\lambda=\pi/2}(t) = \sqrt{\frac{\omega_o}{\omega(t)}}\hat{Q}_{\lambda=\pi/2}\,$. 
From the above results we obtain 
\begin{eqnarray}
\label{eq:finalvariancesrel}
&&  \left(\Delta Q'_{\lambda=0}\right)^{2}(t) = \frac{\omega(t)}{\omega_o}\left(\Delta Q_{\lambda=0}\right)^{2}(t) \nonumber\\
\mbox{and} \:\:\: && \left(\Delta Q'_{\lambda=\pi/2}\right)^{2}(t) = \frac{\omega_o}{\omega(t)}\left(\Delta Q_{\lambda=\pi/2}\right)^{2}(t)\, .
\end{eqnarray}
Thus we have found expressions that allows one to calculate the variance of the instantaneous quadrature operator from the 
the variance of the initial one and vice versa. Similarly, expressions for $R(t)$ as a function of 
$r(t)$ and $\varphi(t)$, as well as for $r(t)$ as a function of $R(t)$ and $\Phi(t)$, 
can be found by substituting Eqs. (\ref{eq:quadratureopvariance}) and (\ref{eq:quadratureopvarianceinst}) in the above equations. 
Finally, notice from Eqs. (\ref{eq:finalvariancesrel}) that 
the product of the variances of the instantaneous quadrature operator with $\lambda=0$ and $\lambda=\pi/2$ is the same as 
for the instantaneous quadrature operator with the same values of lambda.


\section{Continuous interpolation: from sudden to adiabatic}\label{CINT}


In this section we shall study how the SP behaves, once the frequency 
transition has occurred, for different initial and final values of the frequency 
and arbitrary $\epsilon$. 
A comment is in order here: up to the authors knowledge, there is not an analytical 
expression that allows one to calculate exactly the correspondent SP for an arbitrary 
time-dependent frequency of the HO. Hence, some efforts have been made to accomplish this task. 
For instance, a successful approximate analytical expression that works well in the 
sudden limit as well as in the adiabatic one was obtained in the context of analogue models 
\cite{FUJII-2011, FUJII-2015}. However,
this formula is no longer valid in 
the intermediate regime, as correctly stated by these authors.
It is therefore natural to look for an analytical expression for the SP as a function of 
the initial and final frequencies, in addition to $\epsilon$.

\begin{figure}
\centering
\includegraphics*[width=8.5cm]{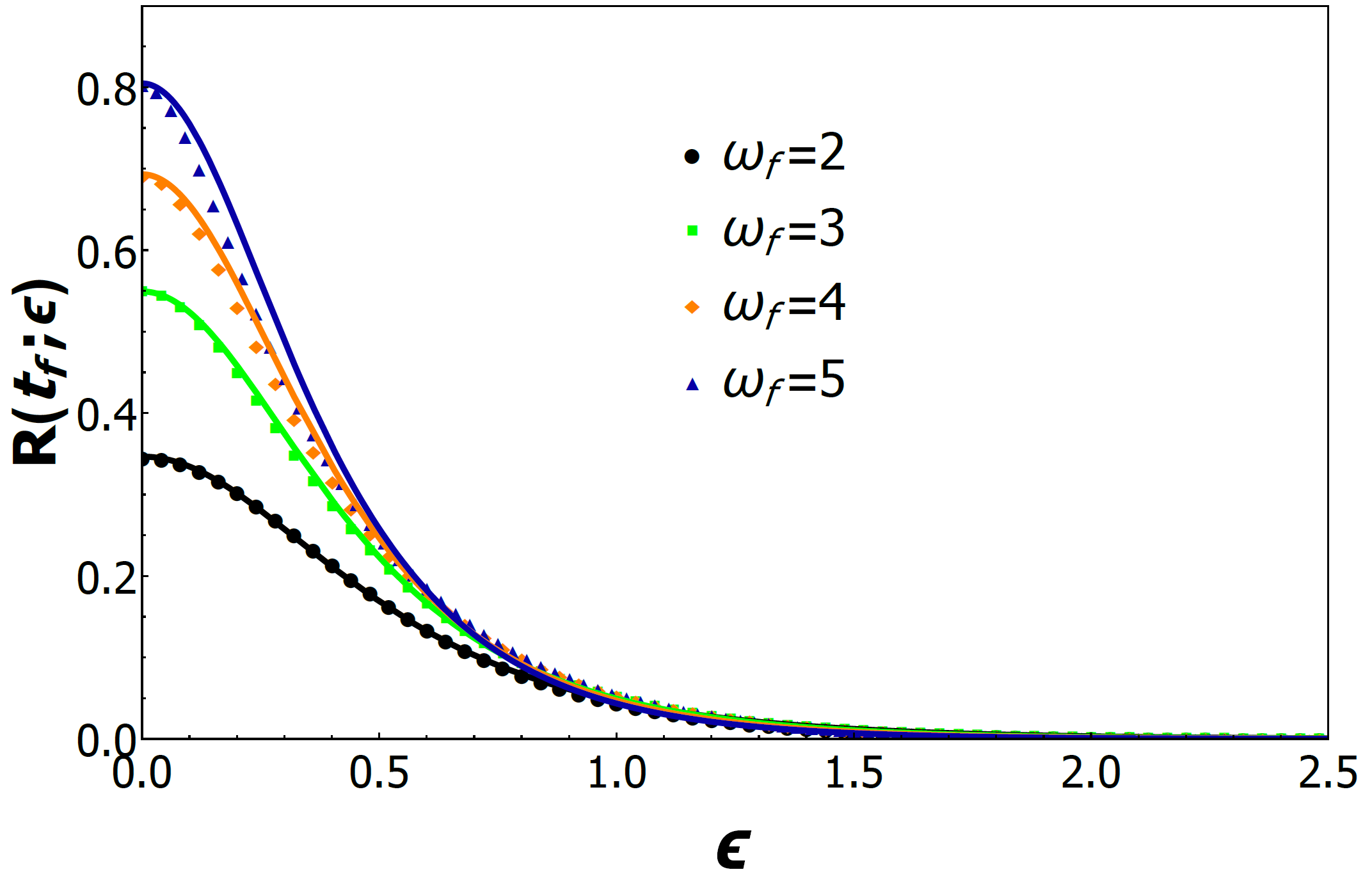}
\caption{$R(t_f;\epsilon)$ as a function of $\epsilon$ for $\omega_{o}=1$ and different values of $\omega_{f}>\omega_{o}$. 
The pointed patterns are numerical results and the solid lines correspond to the fitted results using Eq. (\ref{eq:SPfited}). 
}
\label{fig:spep}
\end{figure}

Let us denote by $R(t_{f};\epsilon)$ the value of $R(t)$ for 
$t> t_0 + 3\epsilon$, \textit{i.e.}, 
after the transition has effectively occurred. 
Initially, we set the initial frequency as $\omega_{o}=1$,   
and calculate $R(t_{f};\epsilon)$ as a function of $\epsilon$, 
for different final values of the frequency with $\omega_{f}>\omega_{o}$. 
These exact (numerical) results correspond to the pointed patterns of Fig. \ref{fig:spep}. 
For instance, the (green) square points in this figure correspond to the case $\omega_{o}=1$ and $\omega_{f}=3$, 
and therefore they show how the final height of the SP, plotted in Fig. \ref{fig:FreqVsTime}(c), 
changes from its largest possible value $R(t_{f};\epsilon=0)=\rho_{f}$, produced by a frequency jump, 
to its smallest possible one $R(t_{f};\epsilon\rightarrow\infty)=0$, obtained for an adiabatic transition. 
As expected, $R(t_{f};\epsilon)$ presents monotonically decreasing behavior as 
the transition becomes slower ($\epsilon$ is increased). 
In other words, the faster the frequency transition occurs, the higher the squeezing. 

Now, we calculate $R(t_{f};\epsilon)$, as a function of $\epsilon$, for a big enough set of 
different initial and final frequencies within the same order of magnitude. 
Using the hyperbolic secant distribution as an Ansatz, 
we interpolated the values of $R(t_{f};\epsilon)$ as a function of 
the initial and final values of the frequency (within a ratio of 10), in addition to $\epsilon$. 
Recalling that $\rho_{f}=\frac{1}{2}\ln{\left(\frac{\omega_{f}}{\omega_{o}}\right)}=\rho(t> t_0 + 3\epsilon)$, 
we obtain the following analytical approximate expression
\begin{equation}
R(t_{f};\epsilon) = \left|\rho_{f}\right|\mbox{sech}\left(2\left[\left|\rho_{f}\right|+1\right]\omega_{min}\epsilon\right) \, ,
\label{eq:SPfited}
\end{equation}
where $\omega_{min}=\mbox{min}\left\{\omega_{o},\omega_{f}\right\}$, 
providing extremely good fittings for $R(t_{f};\epsilon)$. 
Notice that the above equation is valid for $\omega_{f}\leq\omega_{o}$ and $\omega_{f}\geq\omega_{o}$. 
For instance, for the choice made previously, namely, $\omega_{o}=1$ and $\omega_{f}=3\omega_{o}$, 
we should set $\omega_{min}=\omega_{o}$ in Eq. (\ref{eq:SPfited}) to calculate the corresponding fitting.  
Moreover, in Fig. \ref{fig:spep} we plot with solid lines $R(t_{f};\epsilon)$  as a function of $\epsilon$ using 
our fitting given by Eq. (\ref{eq:SPfited})  for the same set of frequencies used in our numerical results (point patterned plots). 
As it can be seen in that figure, the fitting works very well for $\omega_{f}/\omega_{o}\leq5$ 
(error smaller than a few percents) and the agreement gets better the closer the initial and final frequencies are. 
%
\begin{figure}
\centering
\includegraphics*[width=8.5cm]{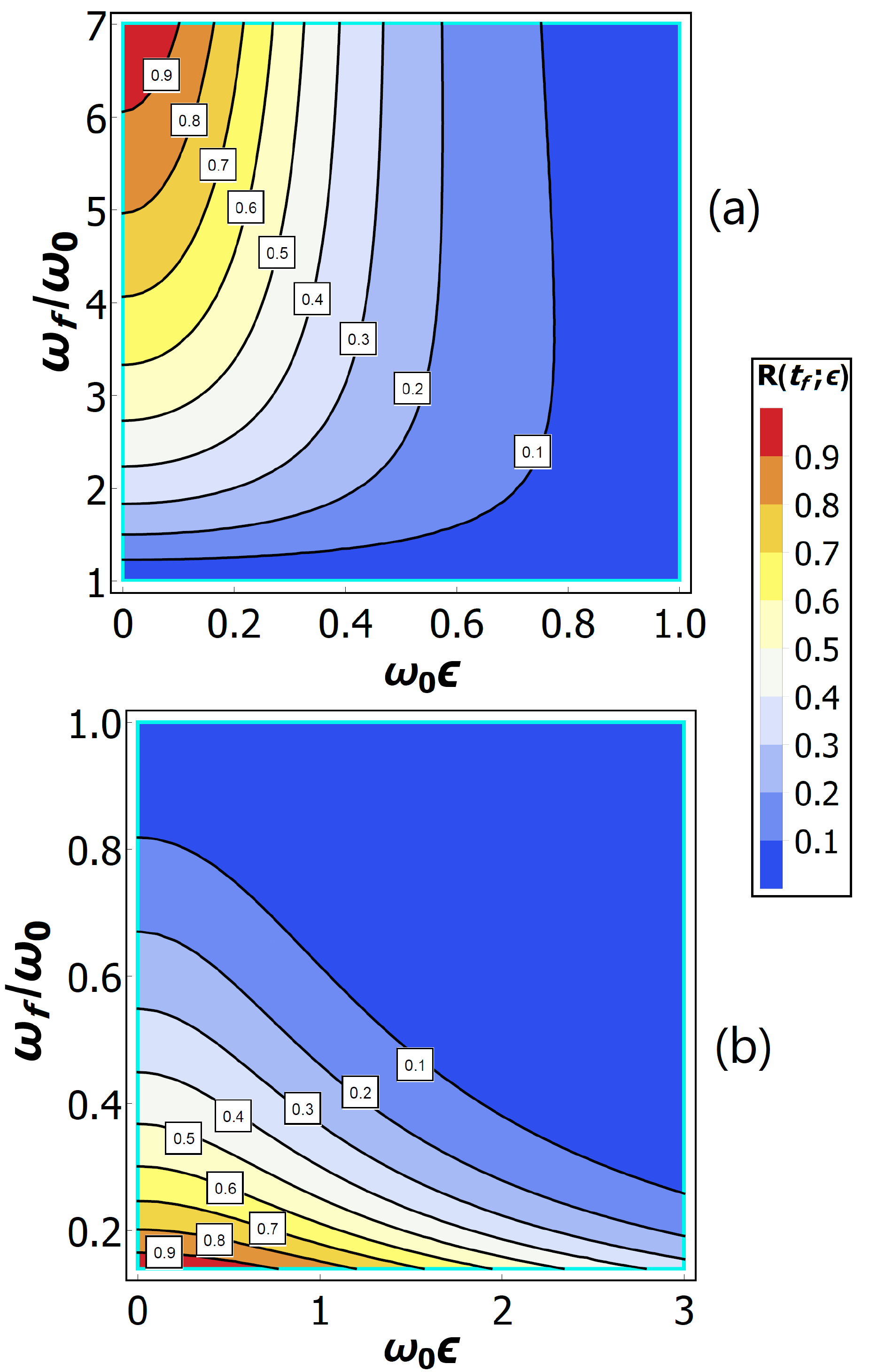}
\caption{Contour plots of Eq. (\ref{eq:SPfited}) for (a) $\omega_{f}/\omega_{o}>1$ and 
(b) $\omega_{f}/\omega_{o}<1$. This graphics enables one to relate the squeezing parameter $R(t_{f};\epsilon)$ with 
relevant parameters of the problem $\omega_{o}$, $\omega_{f}$ and $\epsilon$.}
\label{fig:speplevel}
\end{figure}
Notice that Eq. (\ref{eq:SPfited}) is a precise expression from which we can obtain 
the adiabatic limit for a frequency transition. This can be done by 
taking in such equation $R(t_{f};\epsilon)\rightarrow0$ 
with $\omega_{o}\neq\omega_{f}$, 
resulting in 
\begin{equation}
\frac{\dot{\omega}(t_0)}{\omega_{min}(\omega_{f}-\omega_{o})\left[\left|\rho_{f}\right|+1\right]} << 1 \, ,
\label{eq:Adblimit}
\end{equation}
where it has been used Eq. (\ref{eq:epsiveloci}). 

Now, using Eq. (\ref{eq:SPfited}), in Fig. \ref{fig:speplevel} we present contour 
plots of $R(t_{f};\epsilon)$ as a function of $\omega_{f}/\omega_{o}$ (vertical axis)  
and $\omega_{o}\epsilon$ (horizontal axis), for the two cases:  $\omega_{f}/\omega_{o}>1$  
(panel (a)) and  $\omega_{f}/\omega_{o}<1$ (panel (b)).
Notice that $\omega_{o}$, $\omega_{f}$ and $\epsilon$ are free parameters, 
and the unique restriction is that the frequencies do not differ by a factor greater than 10. 
These contour plots allow one to explore in a more convenient way the space of parameters of the problem. 

To go deeper in the analysis of the above plots, let us initially 
consider the case $\omega_{f}>\omega_{o}$ (Fig. \ref{fig:speplevel}(a)). 
From this plot, it can be seen, for instance, how $R(t_{f};\epsilon)$ 
behaves as we vary $\epsilon$ with $\omega_o$ and $\omega_f$ fixed. 
In other words, we can recover the information contained in Fig. \ref{fig:spep}. 
For our choice made previously, namely, $\omega_{o}=1$ and $\omega_{f}=3\omega_{o}$, 
we can go to Fig. \ref{fig:speplevel}(a), draw a straight horizontal line at $\omega_{f}/\omega_{o}=3$, 
and read the intermediate values of $R(t_{f};\epsilon)$, given by the contour plot, 
as we move along this horizontal line from $\omega_{o}\epsilon = 0$ (sudden limit) to the right. 
It is worth mentioning that, in addition to the choice $\left\{\omega_{f}=3,\omega_{o}=1\right\}$, 
there are an infinity  of other possibilities with the same ratio $\omega_f/\omega_o = 3$. 
An inspection of Fig. \ref{fig:speplevel}(a) also shows that, as the ratio $\omega_f/\omega_o$ tends to $1$, 
the squeezing parameter approaches zero as expected, since in this case there is no change in the frequency at all.

By the same token, we can also see in Fig. \ref{fig:speplevel}(a) how $R(t_{f};\epsilon)$ 
behaves as we vary the ratio $\omega_{f}/\omega_{o}$ for fixed values of $\omega_o\epsilon$. 
It suffices to draw a vertical line and read the intermediate values   
of $R(t_{f};\epsilon)$ as we move upwards along this vertical line from the lowest value $\omega_{f}/\omega_{o} = 1$.
As it can be noted, the maximal values achieved by $R(t_{f};\epsilon)$ (top of the vertical lines) becomes smaller as 
we consider higher values of $\omega_o\epsilon$, as expected, since the greater $\omega_o\epsilon$ is, the more adiabatic the transition will be. 

An advantage of the contour plots shown in  Fig. \ref{fig:speplevel}  is that they provide us 
with a convenient way to search for the values of the relevant parameters that lead to the desired 
values of the squeezing parameter. For instance, suppose we want to know the values of $\omega_f/\omega_o$  
and $\omega_o\epsilon$ such that $R(t_{f};\epsilon) < 0.1$. Looking at Fig. \ref{fig:speplevel} we see 
that the answer to this question is the darkest (blue) region 
limited by the line corresponding to $R(t_{f};\epsilon) = 0.1$. 
 
For completeness, we show in Fig. \ref{fig:speplevel}(b) a contour plot 
totally analogous to that presented in Fig. \ref{fig:speplevel}(a), 
except for the fact that, now, $\omega_f < \omega_o$, so that $\omega_o$ 
is the largest frequency between $\omega_o$ and $\omega_f$. 
Drawing horizontal and vertical lines, a similar analysis to the previous one can also be made. 
As it can be checked, 
for a jump from $\omega_o$ to $\omega_f$ (that is, for $\omega_o\epsilon \rightarrow 0$), 
the value of $R(t_{f};\epsilon)$ will be the same if 
the ratios $\omega_f/\omega_o$ in Fig. \ref{fig:speplevel}(a) and $\omega_o/\omega_f$
in Fig. \ref{fig:speplevel}(b), are the same 
(see the prefactor term in Eq. (\ref{eq:SPfited})). 
However, for finite values of $\omega_o\epsilon$ 
(non-sudden change) this is not true anymore. 
For instance, take the ratios $\omega_f/\omega_o = 5$ in Fig. \ref{fig:speplevel}(a) and 
$\omega_f/\omega_o = 0.2$ (which means $\omega_o/\omega_f = 5$) and look at the value $\omega_o\epsilon = 0.4$. 
In the former case, we see that $0.3 < R(t_{f};\epsilon) < 0.4$, 
while in the latter case $0.8 < R(t_{f};\epsilon) < 0.9$. 
Therefore, starting with $\omega_o$ we will get a larger squeezing for finite values of $\epsilon$ 
if condition $\omega_f < \omega_o \,$ is satisfied instead of $\omega_f > \omega_o\,$. 
Notice that $\omega(t)=0$ corresponds to the free particle situation,  
where the momentum can be infinitely squeezed, 
while the position is totally spread \cite{Ford_2002}. 
Furthermore, in a frequency decreasing (enhancing) the quadrature that starts to be squeezed 
is the one proportional to the momentum (position), while the other quadrature spreads \cite{Averbukh-1994}. 
Our results are consistent with the above description, 
since $R(t_{f};\epsilon)$ increases its value monotonically as the frequency is reduced, 
independently of $\epsilon$.


\section{CONCLUSIONS AND FINAL REMARKS}\label{concl}

In this work we considered a harmonic oscillator (HO) 
undergoing a frequency transition 
modeled by a family of functions 
containing a continuous parameter, denoted by $\epsilon$, 
whose values, when appropriately chosen, 
can tune from a sudden transition
($\epsilon \rightarrow 0$) to an adiabatic one ($\epsilon \rightarrow \infty$). 
The main purpose of our work was to relate squeezing, 
described by the so-called squeezing parameter (SP), 
to adiabaticity, described by $\epsilon$, 
during the time evolution of the transition.
Initially, we fixed the initial ($\omega_o$) and final ($\omega_f$) values of the frequency 
in the transition and studied how squeezing behaves when the frequency rate 
is changed from abrupt to slow. We did that using two descriptions, namely, 
the instantaneous basis, in which the instantaneous hamiltonian is diagonal, 
as well as the initial basis, in which the initial hamiltonian is diagonal. 
The contrast between these two bases descriptions allowed us to identify that
the instantaneous basis affords a direct and intuitive interpretation of the 
adiabatic theorem, while for the initial basis the adiabatic theorem 
must be interpreted accordingly. 
Consequently, we considered a big set of different values of 
$\omega_o$ and $\omega_f$, and studied how 
the final value of the SP, denoted by $R(t_{f};\epsilon)$,
behaves when the frequency rate interpolates continuously 
the sudden and adiabatic limits. 
Using numerical analysis and the hyperbolic secant distribution as an Ansatz, 
we interpolated the values of $R(t_{f};\epsilon)$ as a function of 
$\omega_o$ and $\omega_f$, in addition to $\epsilon$. 
This is how we obtained our main result: an analytical expression that allows one to calculate, 
directly, $R(t_{f};\epsilon)$ as a function of the relevant parameters of the problem, 
namely, $\epsilon$, $\omega_o$ and $\omega_f$ (within the same ratio of 10). 
We showed that our results provides excellent fitting for $R(t_{f};\epsilon)$, 
allowing us to answer the question of how sudden or adiabatic is a given change 
in the frequency of a quantum harmonic oscillator (HO). 
In order to investigate the values of the relevant parameters 
so that $R(t_{f};\epsilon)$ is restricted to desired intervals, 
we made some contour plots for $R(t_{f};\epsilon)$ as a function of 
$\omega_f/\omega_o$ and $\epsilon$. 
An interesting result is that, for a sudden frequency change ($\epsilon = 0$), 
the same  amount of squeezing will be produced if we enhance 
or diminish the frequency of the HO by the same factor. 
However, for non-sudden transitions (finite values of $\epsilon\neq 0$), 
the squeezing will be higher when the frequency is reduced instead of enhanced by the same factor. 
Additionally, we have shown that this result is consistent with obtaining the 
free particle situation as the limiting case of a HO with $\omega(t) \rightarrow 0$.

The model we used for the frequency transition can be considered as 
the initial term in a sequence of transitions to build an arbitrary $\omega(t)$. 
Therefore, we expect our results to be useful in the construction 
of analytical expressions for the SP in HO's with arbitrary frequency modulations.  
Furthermore, the theoretical framework presented in this paper can be generalized to study 
the time evolution of any other initial state, including a thermal one, interesting in 
the realms of shortcuts to adiabaticity and quantum thermodynamics, for instance. 
In such contexts, 
the clarification of squeezing parameters in different basis 
might be interesting to address important questions about the third law of thermodynamics, 
quantum speed limits, or energetic cost of shortcuts.  
We also think that our results could be useful in the study of harmonic traps 
\cite{Grossmann-1995, Uzdin_2013, Hoffmann_2013, Ibarra-2015, schneiter-2020, qvarfort-2020}, 
characterization of adiabaticity in quantum many-body systems \cite{Skelt-2020} 
as well as in problems with coupled HO's \cite{urzua-2019, urzua-2-2019}.


\vskip 0.4cm
\section{Acknowledgments}

The authors acknowledge C. A. D. Zarro, A. L. C. Rego, D. Szilard, Reinaldo F. de Melo e Souza, 
A. Z. Khoury and P.A. Maia Neto for enlightening discussions. 
D. M. T. is also grateful to C. M. H. Willach Galliez and M. Manosalva for their constant support.
The authors thank the Brazilian agency for scientific and technological research CAPES for partial financial support.
This work was partially supported by Conselho Nacional de Desenvolvimento Cient\'{\i}fico e Tecnol\'{o}gico - CNPq, 310365/2018-0 (C.F.).


\bibliographystyle{unsrt}
\bibliography{referencias}

\end{document}